\documentclass[%
 reprint,
superscriptaddress,
preprintnumbers,
nofootinbib,
 amsmath,amssymb,
 aps,
 onecolumn,
 prd,
]{revtex4-2}
\usepackage{graphicx}
\usepackage{dcolumn}
\usepackage{bm}
\usepackage{physics}
\usepackage{xcolor}
\usepackage{comment}
\newcommand{\mpl}{M_{\mathrm{pl}}}

\newcommand{\bp}{\mathbf{p}}

\newcommand{\bk}{\mathbf{k}}
\newcommand{\bz}{\bm{\zeta}}
\newcommand{\bc}{\bm{\chi}}
\newcommand{\md}{\mathrm{d}}
\begin{document}
\preprint{RESCEU-3/23}
\title{Note on the bispectrum and one-loop corrections in single-field inflation with primordial black hole formation}
\author{Jason Kristiano}
\email{jkristiano@resceu.s.u-tokyo.ac.jp}
\affiliation{Research Center for the Early Universe (RESCEU), Graduate School of Science, The University of Tokyo, Tokyo 113-0033, Japan}
\affiliation{Department of Physics, Graduate School of Science, The University of Tokyo, Tokyo 113-0033, Japan \looseness=-1}

\author{Jun'ichi Yokoyama}
\email{yokoyama@resceu.s.u-tokyo.ac.jp}
\affiliation{Kavli Institute for the Physics and Mathematics of the Universe (Kavli IPMU), WPI, UTIAS, The University of Tokyo, Kashiwa, Chiba 277-8568, Japan}
\affiliation{Research Center for the Early Universe (RESCEU), Graduate School of Science, The University of Tokyo, Tokyo 113-0033, Japan}
\affiliation{Department of Physics, Graduate School of Science, The University of Tokyo, Tokyo 113-0033, Japan \looseness=-1}
\affiliation{Trans-Scale Quantum Science Institute, The University of Tokyo, Tokyo 113-0033, Japan}

\date{\today}
\begin{abstract}
Primordial black holes can be formed from the collapse of large-amplitude perturbation on small scales in the early Universe. Such an enhanced spectrum  can be realized by introducing a flat region in the  potential of single-field inflation, which makes the inflaton go into a temporary ultraslow-roll period. In this paper, we calculate the bispectrum of curvature perturbation in such a scenario. We explicitly confirm that bispectrum satisfies Maldacena's theorem. At the end of the ultraslow-roll period, the bispectrum is generated by bulk interaction and field redefinition. At the end of inflation, bispectrum is generated only by bulk interaction. We also calculate the one-loop correction to the power spectrum from the bispectrum, called the source method. We find it consistent with the calculation of the one-loop correction from the second-order expansion of in-in perturbation theory.
\end{abstract}

\maketitle

\section{Introduction}
Despite no observational evidence, primordial black holes (PBHs) have been a research interest \cite{Zel:1967, Hawking:1971ei, Carr:1974nx, Hawking:1974rv} because they are a potential dark matter candidate \cite{Green:2020jor, Carr:2020xqk} and they can explain the origin of binary black holes found by gravitational wave events \cite{Sasaki:2016jop}. The most widely studied formation mechanism of a PBH is the collapse of large-amplitude  quantum fluctuations on small scales generated in single-field inflation. On large scales, quantum fluctuations are tightly constrained by cosmic microwave background (CMB) observation \cite{Planck:2018nkj, Planck:2018jri, Planck:2019kim}. Their power spectrum is almost scale invariant with the amplitude $2.1 \times 10^{-9}$. On small scales, observational constraints are loose enough so it is possible to have a theory of large amplitude of the power spectrum \cite{Nakama:2014vla, Jeong:2014gna, Inomata:2016uip, Nakama:2017qac, Kawasaki:2021yek, Kimura:2021sqz, Wang:2022nml}. Typically $\mathcal{O}(0.01)$ amplitude of power spectrum is needed to produce a significant amount of PBHs \cite{Carr:2009jm, Carr:2020gox}.

In our paper \cite{Kristiano:2022maq}, we pointed out that such a large amplitude of small-scale perturbation can affect prediction on large scales. This is possible because cubic self-interaction between perturbation with long and short wavelengths induces the one-loop correction to the large-scale power spectrum. We considered a PBH formation from an extremely flat region in the potential that induces a temporary ultraslow-roll (USR) period \cite{Kinney:1997ne, Inoue:2001zt, Kinney:2005vj, Martin:2012pe, Motohashi:2017kbs}. In the end, we argued that our result could be generalized into any PBH formation model with a sharp transition of the second slow-roll (SR) parameter. We found that if the small-scale power spectrum reached $\mathcal{O}(0.01)$, the one-loop correction to the large-scale power spectrum would become comparable to its tree-level contribution breaking the perturbativity of the theory. Therefore, we have concluded that PBH formation in single-field inflation is ruled out.

In this paper, we explore the features of this cubic self-interaction. In Sec.~\ref{sec2}, we briefly review the power spectrum of curvature perturbation. In Sec.~\ref{sec3}, we calculate the bispectrum of curvature perturbation generated by cubic self-interaction. We explicitly confirm that it satisfies Maldacena's theorem. In Sec.~\ref{oneloop}, we calculate the one-loop correction to the large-scale power spectrum by two different methods. We find the same result of one-loop correction in both methods. In Sec.~\ref{sec5}, we conclude our paper.

\section{Two-point functions \label{sec2}}

In this section, we briefly review the analytical formula of curvature perturbation in PBH formation \cite{Starobinsky:1992ts, Leach:2001zf, Byrnes:2018txb, Liu:2020oqe, Tasinato:2020vdk, Karam:2022nym, Pi:2022zxs}. We consider a formation model from an extremely flat region in the  potential \cite{Ivanov:1994pa} that leads to a temporary USR motion of the inflaton. The action of canonical inflation is given by
\begin{equation}
S = \frac{1}{2} \int \md^4x \sqrt{-g} \left[ \mpl^2 R - (\partial_\mu \phi)^2 - 2 V(\phi) \right], \label{action}
\end{equation}
where $\mpl$ is reduced Planck scale, $g = \mathrm{det}~g_{\mu\nu}$, $g_{\mu\nu}$ and $R$ are metric tensor and its Ricci scalar. Consider a spatially flat, homogeneous and isotropic background,
\begin{equation}
\md s^2 = -\md t^2 + a^2(t) \md \mathbf{x}^2 = a^2(\tau) (-\md \tau^2 + \md \mathbf{x}^2),
\end{equation}
where $\tau$ is conformal time. Equations of motion for the scale factor $a(t)$ and the homogeneous part of the inflaton $\phi(t)$ are the Friedmann equations,
\begin{equation}
H^2 = \frac{1}{3\mpl^2}\left(\frac{1}{2} \dot{\phi}^2 + V(\phi)\right), ~\dot{H} = - \frac{\dot{\phi}^2}{2 \mpl^2},
\end{equation}
with $H=\dot{a}/{a}$ being the Hubble parameter, and the Klein-Gordon equation,
\begin{equation}
\ddot{\phi} + 3 H \dot{\phi} + \frac{\md V}{\md\phi} = 0. \label{klein}
\end{equation}
Here, a dot denotes time derivative.

\begin{figure}[tbp]
\centering 
\includegraphics[width=0.6\textwidth]{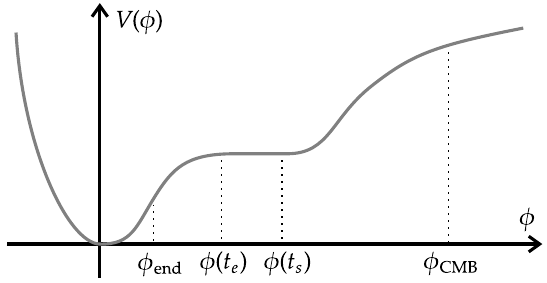}
\caption{\label{fig1} Schematic picture of the inflaton potential realizing PBH formation. When the inflaton is around $\phi_\mathrm{CMB}$, scales probed by CMB observations leave the horizon and it is in the SR regime. It enters an extremely flat region at $t=t_s$ undergoing an USR period. It enters the SR period again at $t=t_e$ until $\phi_\mathrm{end}$, the end of inflation.}
\end{figure}

When CMB-scale fluctuations leave the horizon at around $\phi_\mathrm{CMB}$ (see Fig.~\ref{fig1}), the potential is slightly tilted to realize slow-roll inflation, satisfying 
\begin{equation}
\abs{\frac{\ddot{\phi}}{\dot{\phi} H}} \ll 1, ~\epsilon \equiv - \frac{\dot{H}}{H^2} = \frac{\dot{\phi}^2}{2 \mpl^2 H^2} \ll 1,
\end{equation}
where $\epsilon$ is a SR parameter. In the SR period, $\epsilon$ is approximately constant. Then the inflaton goes through an extremely flat region of the potential, between time $t_s$ to $t_e$, experiencing an USR period. When inflaton enters this region with $\md V / \md \phi \approx 0$, Eq.\ \eqref{klein} becomes $\ddot{\phi} \approx -3H \dot{\phi}$,
so $\dot{\phi} \propto a^{-3}$, which breaks SR approximation \cite{Martin:2012pe}. This makes $\epsilon$ strongly time dependent and extremely small as
\begin{equation}
\epsilon = \frac{\dot{\phi}^2}{2 \mpl^2 H^2}  \propto a^{-6}. \label{epsusr}
\end{equation}
We also define the second SR parameter,
\begin{equation}
\eta \equiv \frac{\dot{\epsilon}}{\epsilon H} = 2 \epsilon +  2 \frac{\ddot{\phi}}{\dot{\phi} H},
\end{equation}
which is approximately constant and very small in SR period $\abs{\eta} \ll 1$, but large in USR period $\eta \approx -6$. 
The latter regime satisfies the condition of the growth of the 
nonconstant mode of perturbation found in \cite{Saito:2008em}, namely, $3-\epsilon+\eta<0$, so that enhanced spectrum is obtained then.

After the USR period, the inflaton enters the SR period again until the end of inflation. In both SR and USR periods, because $\epsilon$ is very small, the scale factor can be approximated as $a = -1/H\tau \propto e^{H t}$.

Small perturbation from the homogeneous part, $\phi(t)$, of the inflaton $\phi(\mathbf{x}, t)$ and metric can be expressed as 
\begin{gather}
\phi(\mathbf{x},t) = \phi(t) + \delta \phi(\mathbf{x},t), \nonumber\\
\md s^2 = -N^2 \mathrm{d}t^2 + \gamma_{ij} (\mathrm{d}x^i + N^i \mathrm{d}t)(\mathrm{d}x^j + N^j \mathrm{d}t),
\end{gather}
where $\gamma_{ij}$ is the three-dimensional metric on slices of constant $t$, $N$ is the lapse function, and $N^i$ is the shift vector. We choose comoving gauge condition
\begin{equation}
\delta \phi(\mathbf{x},t) = 0, ~\gamma_{ij}(\mathbf{x},t) = a^2(t) e^{2\zeta(\mathbf{x},t)} \delta_{ij},
\end{equation}
where $\zeta(\mathbf{x},t)$ is comoving curvature perturbation. Here, tensor perturbation is not relevant. Also, $N$ and $N^i$ are obtained by solving constraint equations.

Expanding the action \eqref{action} up to the second order of the curvature perturbation yields
\begin{equation}
S^{(2)}[\zeta] = M_{\mathrm{pl}}^2 \int \mathrm{d}t ~\mathrm{d}^3x ~a^3 \epsilon  \left[ \dot{\zeta}^2 - \frac{1}{a^2} (\partial_i \zeta)^2  \right].
\label{S2}
\end{equation}
In terms of the Mukhanov-Sasaki variable $v = z \mpl \zeta$, where $z = a \sqrt{2\epsilon}$, the action becomes canonically normalized,
\begin{equation}
S^{(2)}_{\mathrm{can}} [v] = \frac{1}{2} \int \mathrm{d}\tau ~\mathrm{d}^3x \left[ (v')^2 - (\partial_i v)^2 + \frac{z''}{z} v^2 \right],
\end{equation}
where a prime denotes derivative with respect to $\tau$. In momentum space, quantization is performed by promoting the Mukhanov-Sasaki variable as an operator
\begin{equation}
v_\bk (\tau) = M_{\mathrm{pl}} z \zeta_\bk (\tau) =  v_k(\tau) \hat{a}_{\mathbf{k}} + v^*_k (\tau) \hat{a}_{-\mathbf{k}}^\dagger, \nonumber
\end{equation}
where mode function $v_k(\tau)$ approximately satisfies 
\begin{equation}
v_k'' + \left( k^2 - \frac{2}{\tau^2} \right) v_k = 0,
\end{equation}
in both SR and USR regimes, 
and the operators satisfy the commutation relation $[ \hat{a}_{\mathbf{k}}, \hat{a}_{-\mathbf{k'}}^\dagger ] = (2 \pi)^3 \delta(\mathbf{k} + \mathbf{k'})$ under the normalization condition,
\begin{equation}
v_k'^* v_k - v_k' v_k^* = i. \label{normalization}
\end{equation}

The general solution of mode function $v_k(\tau)$ is
\begin{equation}
v_k(\tau) = \frac{\mathcal{A}_k}{\sqrt{2k}} \left( 1-\frac{i}{k\tau} \right) e^{-ik\tau} + \frac{\mathcal{B}_k}{\sqrt{2k}} \left( 1+\frac{i}{k\tau} \right) e^{ik\tau},
\end{equation}
where $\mathcal{A}_k$ and $\mathcal{B}_k$ are determined by boundary conditions or definition of a vacuum state.

At early time, $t \lesssim t_s$, the inflaton was in SR period  with Bunch-Davies initial vacuum.  Then the mode function of the curvature perturbation $\zeta_k = v_k/ z \mpl$ is given by
\begin{equation}
\zeta_k(\tau) = \left( \frac{i H}{2 \mpl \sqrt{\epsilon_{\mathrm{SR}}}} \right)_\star \frac{1}{k^{3/2}} \times\left[ \mathcal{A}_{1,k} e^{-ik\tau} (1+ik\tau) - \mathcal{B}_{1,k} e^{ik\tau} (1-ik\tau) \right], \label{zetasr}
\end{equation}
with the particular choice of $\mathcal{A}_{1,k} = 1$ and $\mathcal{B}_{1,k} = 0$.
Here $\epsilon_{\mathrm{SR}}$ is $\epsilon$ in SR period and subscript $\star$ denotes the value at the horizon crossing epoch $\tau = -1/k$.

At $t_s \lesssim t \lesssim t_e$, the inflaton is in USR period. We define $\tau_s$ and $\tau_e$ as conformal time corresponding to $t_s$ and $t_e$, respectively. The SR parameter $\epsilon$ can be written as $\epsilon(\tau) = \epsilon_{\mathrm{SR}} (\tau / \tau_s)^6$ based on proportionality in \eqref{epsusr}. Therefore, the curvature perturbation becomes,
\begin{equation}
\zeta_k(\tau) = \left( \frac{i H}{2 \mpl \sqrt{\epsilon_{\mathrm{SR}}}} \right)_\star  \left( \frac{\tau_s}{\tau} \right)^3 \frac{1}{k^{3/2}} \times\left[ \mathcal{A}_{2,k} e^{-ik\tau} (1+ik\tau) - \mathcal{B}_{2,k} e^{ik\tau} (1-ik\tau) \right], \label{zetausr}
\end{equation}
where coefficients $\mathcal{A}_{2,k}$ and $\mathcal{B}_{2,k}$ are determined by matching to the SR solution \eqref{zetasr} at the boundary. We consider instantaneous transition from SR to USR, because it is a good approximation to numerical solutions \cite{Karam:2022nym}. Solutions of the coefficients by requiring continuity of $\zeta_k(\tau)$ and $\zeta_k'(\tau)$ at transition $\tau = \tau_s$ are
\begin{gather}
\mathcal{A}_{2,k} = 1 - \frac{3(1 + k^2 \tau_s^2)}{2i k^3 \tau_s^3} \label{coefa2}, \\
\mathcal{B}_{2,k} = - \frac{3(1 + i k \tau_s)^2}{2i k^3 \tau_s^3} e^{-2ik \tau_s}. \label{coefb2}
\end{gather}

At late time, $t \gtrsim t_e$, the inflaton goes back to SR dynamics. The curvature perturbation can be written as
\begin{equation}
\zeta_k(\tau) = \left( \frac{i H}{2 \mpl \sqrt{\epsilon_{\mathrm{SR}}}} \right)_\star  \left( \frac{\tau_s}{\tau_e} \right)^3 \frac{1}{k^{3/2}} \times\left[ \mathcal{A}_{3,k} e^{-ik\tau} (1+ik\tau) - \mathcal{B}_{3,k} e^{ik\tau} (1-ik\tau) \right], \label{zetasr2}
\end{equation}
where coefficients $\mathcal{A}_{3,k}$ and $\mathcal{B}_{3,k}$ are determined by matching to the USR solution \eqref{zetausr} at the boundary. Solutions of the coefficients by requiring continuity of $\zeta_k(\tau)$ and $\zeta_k'(\tau)$ at transition $\tau = \tau_e$ are
\begin{gather}
\mathcal{A}_{3,k} = \frac{-1}{4k^6 \tau_s^3 \tau_e^3} \left\lbrace 9(k\tau_s - i)^2 (k\tau_e + i)^2 e^{2ik(\tau_e - \tau_s)} - [k^2 \tau_s^2 (2 k \tau_s + 3i) + 3i] [k^2 \tau_e^2 (2k \tau_e - 3i) - 3i] \right\rbrace, \label{coefa3} \\
\mathcal{B}_{3,k} = \frac{3}{4k^6 \tau_s^3 \tau_e^3} \left\lbrace e^{-2i k \tau_s} [3 + k^2 \tau_e^2 (3-2i k \tau_e)] (k \tau_s - i)^2 + i e^{-2i k \tau_e} [3i + k^2 \tau_s^2 (2 k \tau_s + 3i)] (k \tau_e - i)^2 \right\rbrace. \label{coefb3}
\end{gather}

The two-point functions of curvature perturbation and power spectrum can be written as
\begin{gather}
\left\langle \zeta_\bk (\tau) \zeta_{\bk'} (\tau) \right\rangle \equiv (2 \pi)^3 \delta^3(\mathbf{k} + \mathbf{k'}) \left\langle \! \left\langle \zeta_\bk (\tau) \zeta_{-\bk} (\tau) \right\rangle \! \right\rangle, \\
\Delta^2_s(k, \tau) \equiv \frac{k^3}{2 \pi^2} \left\langle \! \left\langle \zeta_\bk (\tau) \zeta_{-\bk} (\tau) \right\rangle \! \right\rangle, 
\end{gather}
 the bracket $\langle \cdots \rangle = \bra{0} \cdots \ket{0}$ denotes the vacuum expectation value (VEV), and $\Delta^2_s(k)$ is the power spectrum multiplied by the phase space density. We define $k_s$ and $k_e$ as wavenumbers which cross the horizon at $\tau_s$ and $\tau_e$, respectively. At the end of inflation, $\tau_0~(\rightarrow 0)$, the tree-level power spectrum is
\begin{equation}
\Delta^2_{s(0)}(k, \tau_0) = \frac{k^3}{2 \pi^2} \abs{\zeta_k(\tau_0)}^2 = \left( \frac{H^2}{8 \pi^2 M_{\mathrm{pl}}^2 \epsilon_{\mathrm{SR}}} \right)_\star \left( \frac{k_e}{k_s} \right)^6 \abs{\mathcal{A}_{3,k} - \mathcal{B}_{3,k}}^2, \label{ps0}
\end{equation}
where coefficients $\mathcal{A}_{3,k}$ and $\mathcal{B}_{3,k}$ are given by \eqref{coefa3} and \eqref{coefb3}, respectively.

\begin{figure}[tbp]
\centering 
\includegraphics[width=0.6\textwidth]{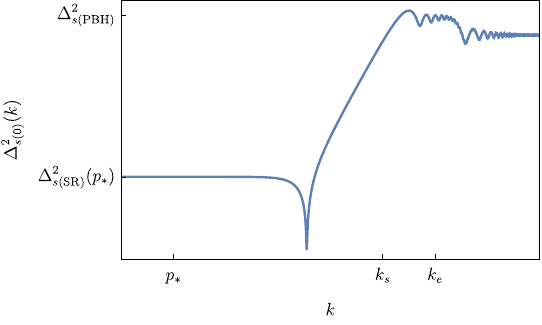}
\caption{\label{fig2} Power spectrum of the curvature perturbation. At CMB scale, $k \ll k_s$, the power spectrum is almost scale invariant. $p_* = 0.05 ~\mathrm{Mpc}^{-1}$ is the pivot scale with amplitude  $\Delta_{s(\mathrm{SR})}^{2}(p_*) = 2.1 \times 10^{-9}$, based on observational result \cite{Planck:2018jri}. At small scale, between $k_s$ and $k_e$, the power spectrum is amplified to typically $\Delta_{s(\mathrm{PBH})}^{2} \sim \mathcal{O}(0.01)$ to form appreciable amount of PBHs.}
\end{figure}

On large scale, the power spectrum approaches an almost scale-invariant limit
\begin{equation}
\Delta_{s(\mathrm{SR})}^{2}(k) \equiv \Delta^2_{s(0)}(k \ll k_s, \tau_0) = \left( \frac{H^2}{8 \pi^2 M_{\mathrm{pl}}^2 \epsilon_{\mathrm{SR}}} \right)_\star, \label{pssr}
\end{equation}
with a small wavenumber dependence due to the horizon crossing condition manifested in the spectral tilt
\begin{equation}
n_s - 1 = \frac{\md \log \Delta_{s(\mathrm{SR})}^{2}}{\md \log k} = - 2 \epsilon_{\mathrm{SR}} - \eta_{\mathrm{SR}},
\end{equation}
where $\eta_{\mathrm{SR}}$ is $\eta$ in SR period. This large-scale limit must be consistent with CMB observation. On small scale with larger wavenumber, $k_s \lesssim k \lesssim k_e$, the power spectrum is oscillating around,
\begin{equation}
\Delta_{s(\mathrm{PBH})}^{2} \approx \Delta_{s(\mathrm{SR})}^{2}(k_s) \left( \frac{k_e}{k_s} \right)^6,
\end{equation}
whose high-density peak may collapse into PBHs. It is amplified by factor $(k_e/k_s)^6$ compared to the CMB-scale power spectrum. Plot of the typical power spectrum is shown in Fig.~\ref{fig2}.

\section{Three-point functions \label{sec3}}

Three-point functions is generated by cubic self-interaction. Expanding \eqref{action} to third-order of $\zeta$ yields the interaction action \cite{Maldacena:2002vr}
\begin{equation}
S^{(3)}[\zeta] = S_{\mathrm{bulk}}[\zeta] + S_\mathrm{B}[\zeta] + \mpl^2  \int \md t ~\md^3 x ~2 f(\zeta) \left( \frac{\delta L}{\delta \zeta} \right)_1, \label{S3}
\end{equation}
where the explicit form will be given shortly. The bulk interaction $S_{\mathrm{bulk}}[\zeta]$ reads 
\begin{equation}
S_{\mathrm{bulk}}[\zeta] = \mpl^2  \int \md t ~\md^3 x ~a^3 \left[ \epsilon^2  \dot{\zeta}^2 \zeta + \frac{1}{a^2} \epsilon^2 (\partial_i \zeta)^2 \zeta - 2 \epsilon  \dot{\zeta} \partial_i \zeta \partial_i \chi -\frac{1}{2} \epsilon^3 \dot{\zeta}^2 \zeta + \frac{1}{2} \epsilon  \zeta (\partial_i \partial_j \chi)^2 + \frac{1}{2} \epsilon \dot{\eta} \dot{\zeta} \zeta^2 \right], \label{sbulk}
\end{equation}
where $\chi = \epsilon \partial^{-2} \dot{\zeta}$. The boundary interaction  $S_\mathrm{B}[\zeta]$ reads \cite{Arroja:2011yj, Burrage:2011hd}
\begin{align}
S_\mathrm{B}[\zeta] = \mpl^2  \int \md t ~\md^3 x ~\frac{\md}{\md t} & \left[ - 9a^3 H \zeta^3 + \frac{a}{H} \zeta (\partial_i \zeta)^2 - \frac{1}{4aH^3} (\partial_i \zeta)^2 \partial^2 \zeta - \frac{a \epsilon}{H} \zeta (\partial_i \zeta)^2 \right. \label{sb} \\
& \left. + \frac{a}{2 H^2} \zeta (\partial_i \partial_j \zeta \partial_i \partial_j \chi - \partial^2 \zeta \partial^2 \chi) - \frac{a^3}{2 H^2} \zeta (\partial_i \partial_j \chi \partial_i \partial_j \chi - \partial^2 \chi \partial^2 \chi) - \frac{\epsilon a^3}{H} \zeta \dot{\zeta}^2 - \frac{\eta a^3}{2} \zeta^2 \partial^2 \chi  \right], \nonumber 
\end{align}
where total spatial derivatives are omitted. Boundary interactions without $\dot{\zeta}$ are unimportant because they will not contribute to the correlation of $\zeta$. The last term is interaction proportional to the equation of motion in the lowest order,
\begin{equation}
\left( \frac{\delta L}{\delta \zeta} \right)_1 = \frac{\md }{\md t} (\epsilon a^3 \dot{\zeta}) - \epsilon a \partial^2 \zeta.
\end{equation}
The function $f(\zeta)$ is explicitly given by
\begin{equation}
f(\zeta) = \frac{\eta}{4} \zeta^2 + \frac{\dot{\zeta}}{H} \zeta + \frac{1}{4a^2 H^2} [-(\partial_i \zeta)^2 + \partial^{-2}\partial_i \partial_j (\partial_i \zeta \partial_j \zeta)] + \frac{1}{2H}[\partial_i \zeta \partial_i \chi - \partial^{-2} \partial_i \partial_j (\partial_i \zeta \partial_j \chi)]. \label{redef}
\end{equation}

Performing field redefinition $\zeta = \bz + f(\bz)$ generates a third-order terms from the second-order action \eqref{S2} as
\begin{align}
S^{(2)}[\zeta] = S^{(2)}[\bz] + \int  \md t ~\md^3 x & \left[ (-2) f(\bz) \left( \frac{\delta L}{\delta \bz} \right)_1 - \frac{\md}{\md t} \left( \frac{a}{2 H^2} \bz (\partial_i \partial_j \bz \partial_i \partial_j \bc - \partial^2 \bz \partial^2 \bc) \right. \right. \nonumber\\
& \left. \left. - \frac{a^3}{2 H^2} \bz (\partial_i \partial_j \bc \partial_i \partial_j \bc - \partial^2 \bc \partial^2 \bc) - \frac{\epsilon a^3}{H} \bz \dot{\bz}^2 - \frac{\eta a^3}{2} \bz^2 \partial^2 \bc \right) \right],
\end{align}
where $\bc = \epsilon \partial^{-2} \dot{\bz}$. Such additional third-order action cancels the last term in \eqref{S3} and all the boundary interactions including $\dot{\zeta}$ in \eqref{sb}, so the total action is \footnote{We understand that there should be additional fourth-order action of $\bz$, which arises from substituting field redefinition to the cubic-order action. This should not be an issue in calculating bispectrum. However, such quartic self-interaction can affect one-loop correction to the power spectrum. Fortunately, the fourth-order action of $\bz$ has been calculated by \cite{Jarnhus:2007ia, Arroja:2008ga} and it is expected to generates independent contributions to the one-loop correction than those induced by the cubic self-interaction because it involves higher-order SR parameter. We will calculate one-loop correction induced by the cubic self-interaction in \S~\ref{oneloop}.} \cite{Maldacena:2002vr, Arroja:2011yj, Burrage:2011hd}
\begin{equation}
S^{(2)}[\zeta] + S^{(3)}[\zeta] = S^{(2)}[\bz] + S_{\mathrm{bulk}}[\bz]. \label{sbz}
\end{equation}
In terms of $\bz$, the total action is simply given by the second-order action \eqref{S2} and bulk interaction \eqref{sbulk}. The interaction Hamiltonian in terms of $\bz$ is
\begin{equation}
H_{\mathrm{int}} = - \mpl^2 \int \md^3x ~a^2 \left[ \epsilon^2 (\bz')^2 \bz + \epsilon^2 (\partial_i \bz)^2 \bz - 2\epsilon^2 \bz' \partial_i \bz \partial_i \partial^{-2} \bz' -\frac{1}{2} \epsilon^3 (\bz')^2 \bz + \frac{1}{2} \epsilon^3  \bz (\partial_i \partial_j \partial^{-2} \bz')^2 + \frac{1}{2} \epsilon \eta' \bz' \bz^2 \right]. \label{hint}
\end{equation}
In the standard SR inflation without PBH formation, the first three terms and last three terms of \eqref{hint} have coupling $\mathcal{O}(\epsilon^2)$ and $\mathcal{O}(\epsilon^3)$, respectively. However, for inflation model with PBH formation, everything is the same except the last term of \eqref{hint} becomes $\mathcal{O}(\epsilon)$ because $\eta$ has $\mathcal{O}(1)$ transition \cite{Namjoo:2012aa, Cai:2016ngx, Chen:2013eea, Cai:2018dkf, Davies:2021loj}. Therefore, the leading interaction is
\begin{equation}
H_{\mathrm{int}}(\tau) = - \frac{1}{2} \mpl^2 \int \md^3x ~a^2 \epsilon \eta' \bz' \bz^2 . \label{hintl}
\end{equation}

Three-point functions of $\zeta$ can be written schematically as
\begin{equation}
\langle \zeta_{\bk_1}(\tau) \zeta_{\bk_2}(\tau) \zeta_{\bk_3}(\tau) \rangle = \langle \bz_{\bk_1}(\tau) \bz_{\bk_2}(\tau) \bz_{\bk_3}(\tau) \rangle + \mathrm{redefinition ~terms},
\end{equation}
where the first term is generated by bulk interaction \eqref{hint} and the second term is boundary contribution at $\tau$ from field redefinition. Higher-order correction to the expectation value of an operator $\mathcal{O}(\tau)$ is calculated by the in-in perturbation theory
\begin{equation}
\langle \mathcal{O(\tau)} \rangle =  \left\langle \left[ \bar{\mathrm{T}} \exp \left( i \int_{-\infty}^{\tau} \mathrm{d}\tau' H_{\mathrm{int}}(\tau') \right) \right] \mathcal{\hat{O}} (\tau) \left[ \mathrm{T} \exp \left( -i \int_{-\infty}^{\tau} \mathrm{d\tau'} H_{\mathrm{int}}(\tau') \right) \right] \right\rangle, \label{inin}
\end{equation}
where $\mathrm{T}$ and $\bar{\mathrm{T}}$ denote time and antitime ordering. For three-point functions, the operator is $\bz_{\bk_1}(\tau) \bz_{\bk_2}(\tau) \bz_{\bk_3}(\tau)$. The first-order expansion of \eqref{inin} reads
\begin{equation}
\langle \mathcal{O(\tau)} \rangle = 2 \int_{-\infty}^{\tau} \md\tau_1 ~\mathrm{Im} \left\langle \mathcal{\hat{O}} (\tau) H_{\mathrm{int}}(\tau_1) \right\rangle. \label{inin1}
\end{equation}
The bispectrum $\left\langle \! \left\langle \zeta_{\bk_1}(\tau) \zeta_{\bk_2}(\tau) \zeta_{\bk_3}(\tau) \right\rangle \! \right\rangle$ is defined as
\begin{equation}
\langle \zeta_{\bk_1}(\tau) \zeta_{\bk_2}(\tau) \zeta_{\bk_3}(\tau) \rangle = (2\pi)^3 \delta(\bk_1 + \bk_2 + \bk_3) \left\langle \! \left\langle \zeta_{\bk_1}(\tau) \zeta_{\bk_2}(\tau) \zeta_{\bk_3}(\tau) \right\rangle \! \right\rangle.
\end{equation}
The squeezed limit, when one of wavenumbers in the bispectrum is very small, the bispectrum satisfies Maldacena's theorem \footnote{Maldacena's theorem is valid for attractor single-clock inflation. When a non-attractor period exists, it is generalized as described in \cite{Bravo:2017wyw, Finelli:2017fml}. As an example, generalized Maldacena's theorem in the USR period is
\begin{equation}
\lim_{k_1 \rightarrow 0} \langle \! \langle \zeta_{\bk_1}(\tau) \zeta_{\bk_2}(\tau) \zeta_{-\bk_2}(\tau) \rangle \! \rangle = \frac{1}{6H^2} \dot{P}_s(k_1, \tau) \left[ (n_s(k_2, \tau) - 1) H P_s(k_2, \tau) + \dot{P}_s(k_2, \tau) \right] + (1 - n_s(k_2, \tau)) P_s(k_1, \tau) P_s(k_2, \tau),
\end{equation}
where $P_s(k, \tau) \equiv \left\langle \! \left\langle \zeta_{\bk}(\tau) \zeta_{-\bk}(\tau) \right\rangle \! \right\rangle$. In our context, for perturbation with $p \ll k_s$, $\dot{P}_s(p, \tau_e) \ll H P_s(p, \tau_e)$ as we can check from \eqref{zetausr}, so the consistency condition reduces to the standard one \cite{Riotto:2023hoz}. This is also expected based on \cite{Passaglia:2018ixg}.} \cite{Maldacena:2002vr}
\begin{align}
\lim_{k_1 \rightarrow 0} \left\langle \! \left\langle \zeta_{\bk_1}(\tau) \zeta_{\bk_2}(\tau) \zeta_{-\bk_2}(\tau) \right\rangle \! \right\rangle & = - (n_s(k_2, \tau) - 1) \left\langle \! \left\langle \zeta_{\bk_2}(\tau) \zeta_{-\bk_2}(\tau) \right\rangle \! \right\rangle \left\langle \! \left\langle \zeta_{\bk_1}(\tau) \zeta_{-\bk_1}(\tau) \right\rangle \! \right\rangle \nonumber\\
& = - (n_s(k_2, \tau) - 1) \abs{\zeta_{k_2}(\tau)}^2 \abs{\zeta_{k_1}(\tau)}^2,
\end{align}
where we define
\begin{equation}
n_s(k, \tau) - 1 = \frac{\md \log \Delta_s^2(k, \tau)}{\md \log k}.
\end{equation}
We emphasize that Maldacena's theorem is satisfied by the bispectrum of $\zeta$, not $\bz$. We now calculate the bispectrum at four different epochs.

\subsection{Slightly after the end of USR period \label{usr1}}
\begin{figure}[tbp]
\centering 
\includegraphics[width=0.7\textwidth]{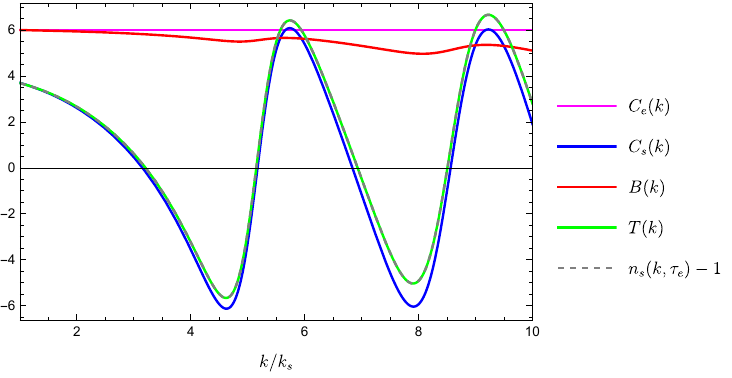}
\caption{\label{fig3} Plot of $C_e(k)$, $C_s(k)$, $B(k)$, $T(k)$, and $n_s(k, \tau_e) - 1$. We choose $k_e/k_s = 10$ just for illustrative purposes.}
\end{figure}

In this subsection, we calculate bispectrum at a time slightly after the end of the USR period. Note that every time $\tau_e$ is written in this subsection implicitly means it is evaluated at $\tau_e^+$ when the inflaton is already in the second SR period. The three-point functions of $\bz$ is given by
\begin{equation}
\langle \bz_{\bk_1}(\tau_e) \bz_{\bk_2}(\tau_e) \bz_{\bk_3}(\tau_e) \rangle = 2 \int_{-\infty}^{\tau_e} \md\tau_1 ~\mathrm{Im} \left\langle \bz_{\bk_1}(\tau) \bz_{\bk_2}(\tau) \bz_{\bk_3}(\tau) H_{\mathrm{int}}(\tau_1) \right\rangle.
\end{equation}
Substituting in \eqref{hintl} yields the bispectrum,
\begin{equation}
\left\langle \! \left\langle \bz_{\bk_1}(\tau_e) \bz_{\bk_2}(\tau_e) \bz_{\bk_3}(\tau_e) \right\rangle \! \right\rangle = -2 \mpl^2 \int_{-\infty}^{\tau_e} \md\tau_1 \epsilon(\tau_1) \eta'(\tau_1) a^2(\tau_1)  \mathrm{Im} \left[ \zeta_{k_1}(\tau_e) \zeta_{k_2}(\tau_e) \zeta_{k_3}(\tau_e) \zeta_{\tilde{k}_1}^*(\tau_1) \zeta_{\tilde{k}_2}^*(\tau_1) \zeta_{\tilde{k}_3}'^*(\tau_1) \right] + \mathrm{perm}. 
\end{equation}
Here and hereafter, $+ \mathrm{perm}$ means summation over a cyclic permutation of suffices with a tilde in the preceeding expression. To evaluate the time integral, we remind that $\eta$ is almost constant in both SR and USR periods, so $\eta'(\tau) \approx 0$ except for sharp transitions around $\tau=\tau_s$ and $\tau=\tau_e$. Therefore, $\eta'(\tau)$ can be written as
\begin{equation}
\eta'(\tau) = \Delta\eta [ -\delta(\tau - \tau_s) + \delta(\tau - \tau_e) ], \label{etadot}
\end{equation}
where $\Delta\eta \approx 6$. After evaluating the time integral, the bispectrum becomes,
\begin{align}
\left\langle \! \left\langle \bz_{\bk_1}(\tau_e) \bz_{\bk_2}(\tau_e) \bz_{\bk_3}(\tau_e) \right\rangle \! \right\rangle = & -2 \mpl^2 \epsilon(\tau_e)  a^2(\tau_e) \Delta\eta ~ \zeta_{k_1}(\tau_e) \zeta_{\tilde{k}_1}^*(\tau_e) \zeta_{k_2}(\tau_e) \zeta_{\tilde{k}_2}^*(\tau_e) \mathrm{Im}(\zeta_{k_3}(\tau_e) \zeta_{\tilde{k}_3}'^*(\tau_e))  \nonumber\\
& + 2 \mpl^2 \epsilon(\tau_s)  a^2(\tau_s) \Delta\eta ~\mathrm{Im} \left[ \zeta_{k_1}(\tau_e) \zeta_{k_2}(\tau_e) \zeta_{k_3}(\tau_e) \zeta_{\tilde{k}_1}^*(\tau_s) \zeta_{\tilde{k}_2}^*(\tau_s) \zeta_{\tilde{k}_3}'^*(\tau_s) \right] + \mathrm{perm}. \label{bispeclong}
\end{align}
From the normalization condition \eqref{normalization}, we can obtain,
\begin{equation}
\mathrm{Im}(\zeta_{k}(\tau_e) \zeta_{k}'^*(\tau_e)) = \frac{1}{4 \mpl^2 \epsilon(\tau_e) a^2(\tau_e) }. \label{imzeta}
\end{equation}
Substituting it into \eqref{bispeclong}, we find the squeezed limit of the bispectrum as
\begin{equation}
\left\langle \! \left\langle \bz_{\bk_1}(\tau_e) \bz_{\bk_2}(\tau_e) \bz_{-\bk_2}(\tau_e) \right\rangle \! \right\rangle = - \left\lbrace \Delta\eta - 4 \Delta\eta \mpl^2 \epsilon(\tau_s) a^2(\tau_s) \mathrm{Im} \left[ \frac{\zeta_{k_2}^2(\tau_e)}{\abs{\zeta_{k_2}(\tau_e)}^2} \zeta_{k_2}^*(\tau_s) \zeta_{k_2}'^*(\tau_s) \right] \right\rbrace  \abs{\zeta_{k_1}(\tau_e)}^2 \abs{\zeta_{k_2}(\tau_e)}^2.
\end{equation}
We define a coefficient $C(k) = C_e(k) + C_s(k)$, where
\begin{equation}
C_e(k) = \Delta\eta, ~C_s(k) =  - 4 \Delta\eta \mpl^2 \epsilon(\tau_s) a^2(\tau_s) \mathrm{Im} \left[ \frac{\zeta_{k}^2(\tau_e)}{\abs{\zeta_{k}(\tau_e)}^2} \zeta_{k}^*(\tau_s) \zeta_{k}'^*(\tau_s) \right]. \label{cecs}
\end{equation}
Plot of $C_e(k)$ and $C_s(k)$ are shown in Fig. \ref{fig3}. For comparison, we also plot $n_s(k, \tau_e) - 1$ directly from the mode function \eqref{zetausr}. We can see that $C_s(k)$ almost overlaps with $n_s(k, \tau_e) - 1$, but they are not precisely equal. Clearly, the coefficient $C(k) \neq n_s(k, \tau_e) - 1$, implies that the bispectrum of $\bz$ does not satisfy Maldacena's theorem.

The bispectrum of $\zeta$ is given by the bispectrum of $\bz$ plus contribution from field redefinition. Because $\eta(\tau_e) \approx 0$, the leading field redefinition \eqref{redef} is $\dot{\bz} \bz/H$. Noting that $a(\tau_e) H = k_e$, the bispectrum of $\zeta$ reads,
\begin{align}
\left\langle \! \left\langle \zeta_{\bk_1}(\tau_e) \zeta_{\bk_2}(\tau_e) \zeta_{\bk_3}(\tau_e) \right\rangle \! \right\rangle = & \left\langle \! \left\langle \bz_{\bk_1}(\tau_e) \bz_{\bk_2}(\tau_e) \bz_{\bk_3}(\tau_e) \right\rangle \! \right\rangle +  \nonumber\\
& k_e^{-1} \left[ \langle \! \langle \bz_{\bk_2}'(\tau_e) \bz_{-\bk_2}(\tau_e) \rangle \! \rangle \left\langle \! \left\langle \bz_{\bk_3}(\tau_e) \bz_{-\bk_3}(\tau_e) \right\rangle \! \right\rangle + \left\langle \! \left\langle \bz_{\bk_2}(\tau_e) \bz_{-\bk_2}(\tau_e) \right\rangle \! \right\rangle \langle \! \langle \bz_{\bk_3}'(\tau_e) \bz_{-\bk_3}(\tau_e) \rangle \! \rangle \right. \nonumber\\
& \left. + \left\langle \! \left\langle \bz_{\bk_1}(\tau_e) \bz_{-\bk_1}(\tau_e) \right\rangle \! \right\rangle \langle \! \langle \bz_{\bk_3}'(\tau_e) \bz_{-\bk_3}(\tau_e) \rangle \! \rangle + \langle \! \langle \bz_{\bk_1}(\tau_e) \bz_{-\bk_1}'(\tau_e) \rangle \! \rangle \left\langle \! \left\langle \bz_{\bk_3}(\tau_e) \bz_{-\bk_3}(\tau_e) \right\rangle \! \right\rangle \right.  \nonumber\\
& \left. + \left\langle \! \left\langle \bz_{\bk_1}(\tau_e) \bz_{-\bk_1}(\tau_e) \right\rangle \! \right\rangle \langle \! \langle \bz_{\bk_2}(\tau_e) \bz_{-\bk_2}'(\tau_e) \rangle \! \rangle + \langle \! \langle \bz_{\bk_1}(\tau_e) \bz_{-\bk_1}'(\tau_e) \rangle \! \rangle \left\langle \! \left\langle \bz_{\bk_2}(\tau_e) \bz_{-\bk_2}(\tau_e) \right\rangle \! \right\rangle \right].
\end{align}
The leading terms in the squeezed limit, $k_1 \rightarrow 0$, are
\begin{align}
\left\langle \! \left\langle \zeta_{\bk_1}(\tau_e) \zeta_{\bk_2}(\tau_e) \zeta_{-\bk_2}(\tau_e) \right\rangle \! \right\rangle = & \left\langle \! \left\langle \bz_{\bk_1}(\tau_e) \bz_{\bk_2}(\tau_e) \bz_{-\bk_2}(\tau_e) \right\rangle \! \right\rangle \nonumber\\
& + k_e^{-1} \left\langle \! \left\langle \bz_{\bk_1}(\tau_e) \bz_{-\bk_1}(\tau_e) \right\rangle \! \right\rangle \left[ \langle \! \langle \bz_{\bk_2}'(\tau_e) \bz_{-\bk_2}(\tau_e) \rangle \! \rangle + \langle \! \langle \bz_{\bk_2}(\tau_e) \bz_{-\bk_2}'(\tau_e) \rangle \! \rangle \right],
\end{align}
and more explicitly
\begin{align}
\left\langle \! \left\langle \zeta_{\bk_1}(\tau_e) \zeta_{\bk_2}(\tau_e) \zeta_{-\bk_2}(\tau_e) \right\rangle \! \right\rangle = - &\left\lbrace \Delta\eta - 4 \Delta\eta \mpl^2 \epsilon(\tau_s) a^2(\tau_s) \mathrm{Im} \left[ \frac{\zeta_{k_2}^2(\tau_e)}{\abs{\zeta_{k_2}(\tau_e)}^2} \zeta_{k_2}^*(\tau_s) \zeta_{k_2}'^*(\tau_s) \right] \right.  \nonumber\\
& \left. - 2 \mathrm{Re} \left[ \frac{\zeta_{k_2}(\tau_e) \zeta_{k_2}'^*(\tau_e)}{k_e \abs{\zeta_{k_2}(\tau_e)}^2} \right] \right\rbrace \abs{\zeta_{k_1}(\tau_e)}^2 \abs{\zeta_{k_2}(\tau_e)}^2. \label{bispec-}
\end{align}
We define contribution from field redefinition as
\begin{equation}
B(k) = 2 \mathrm{Re} \left[ \frac{\zeta_{k}(\tau_e) \zeta_{k}'^*(\tau_e)}{k_e \abs{\zeta_{k}(\tau_e)}^2} \right], \label{bk}
\end{equation}
so the total coefficient is $T(k) = C(k) - B(k)$. The plot of $B(k)$ and $T(k)$ are shown in Fig. \ref{fig3}. We can see that $T(k)$ precisely equals to $n_s(k, \tau_e) - 1$, which confirms Maldacena's theorem. Equality $T(k) = n_s(k, \tau_e) - 1$ holds exactly, which one can confirm by substituting mode function \eqref{zetausr} to the explicit form of $T(k)$.

\subsection{Slightly before the end of USR period \label{usr2}}
In this subsection, we calculate bispectrum at a time slightly before the end of the USR period. Note that every time $\tau_e$ is written in this subsection implicitly means it is evaluated at $\tau_e^-$ when the inflaton is still in the USR period. At this time, there is still no sharp transition at $\tau_e$, so the bispectrum of $\bz$ is
\begin{equation}
\left\langle \! \left\langle \bz_{\bk_1}(\tau_e) \bz_{\bk_2}(\tau_e) \bz_{\bk_3}(\tau_e) \right\rangle \! \right\rangle = 2 \mpl^2 \epsilon(\tau_s)  a^2(\tau_s) \Delta\eta ~\mathrm{Im} \left[ \zeta_{k_1}(\tau_e) \zeta_{k_2}(\tau_e) \zeta_{k_3}(\tau_e) \zeta_{\tilde{k}_1}^*(\tau_s) \zeta_{\tilde{k}_2}^*(\tau_s) \zeta_{\tilde{k}_3}'^*(\tau_s) \right] + \mathrm{perm}. 
\end{equation}
Because $\eta(\tau_e) \approx -6$, the leading field redefinitions \eqref{redef} are $\eta \bz^2/4$ and $\dot{\bz} \bz/H$. In the squeezed limit, the bispectrum of $\zeta$ is
\begin{align}
\left\langle \! \left\langle \zeta_{\bk_1}(\tau_e) \zeta_{\bk_2}(\tau_e) \zeta_{-\bk_2}(\tau_e) \right\rangle \! \right\rangle = & \left\langle \! \left\langle \bz_{\bk_1}(\tau_e) \bz_{\bk_2}(\tau_e) \bz_{-\bk_2}(\tau_e) \right\rangle \! \right\rangle + \frac{\eta(\tau_e)}{4} 4 \left\langle \! \left\langle \bz_{\bk_1}(\tau_e) \bz_{-\bk_1}(\tau_e) \right\rangle \! \right\rangle \left\langle \! \left\langle \bz_{\bk_2}(\tau_e) \bz_{-\bk_2}(\tau_e) \right\rangle \! \right\rangle \nonumber\\
& + k_e^{-1} \left\langle \! \left\langle \bz_{\bk_1}(\tau_e) \bz_{-\bk_1}(\tau_e) \right\rangle \! \right\rangle \left( \langle \! \langle \bz_{\bk_2}'(\tau_e) \bz_{-\bk_2}(\tau_e) \rangle \! \rangle + \langle \! \langle \bz_{\bk_2}(\tau_e) \bz_{-\bk_2}'(\tau_e) \rangle \! \rangle \right).
\end{align}
More explicitly, it can be written as
\begin{align}
\left\langle \! \left\langle \zeta_{\bk_1}(\tau_e) \zeta_{\bk_2}(\tau_e) \zeta_{-\bk_2}(\tau_e) \right\rangle \! \right\rangle = - &\left\lbrace - 4 \Delta\eta \mpl^2 \epsilon(\tau_s) a^2(\tau_s) \mathrm{Im} \left[ \frac{\zeta_{k_2}^2(\tau_e)}{\abs{\zeta_{k_2}(\tau_e)}^2} \zeta_{k_2}^*(\tau_s) \zeta_{k_2}'^*(\tau_s) \right] \right.  \nonumber\\
& \left. - 2 \mathrm{Re} \left[ \frac{\zeta_{k_2}(\tau_e) \zeta_{k_2}'^*(\tau_e)}{k_e \abs{\zeta_{k_2}(\tau_e)}^2} \right] + \Delta\eta \right\rbrace \abs{\zeta_{k_1}(\tau_e)}^2 \abs{\zeta_{k_2}(\tau_e)}^2. \label{bispec+}
\end{align}

\subsection{Exactly at the end of USR period \label{usr3}}
In this subsection, we calculate the bispectrum at exactly the end of the USR period. The bispectrum of $\bz$ is
\begin{align}
\left\langle \! \left\langle \bz_{\bk_1}(\tau_e) \bz_{\bk_2}(\tau_e) \bz_{\bk_3}(\tau_e) \right\rangle \! \right\rangle = & -2 \mpl^2 \epsilon(\tau_e)  a^2(\tau_e) \frac{1}{2} \Delta\eta ~\zeta_{k_1}(\tau_e) \zeta_{\tilde{k}_1}^*(\tau_e) \zeta_{k_2}(\tau_e) \zeta_{\tilde{k}_2}^*(\tau_e) \mathrm{Im}(\zeta_{k_3}(\tau_e) \zeta_{\tilde{k}_3}'^*(\tau_e)) \nonumber\\
 & + 2 \mpl^2 \epsilon(\tau_s)  a^2(\tau_s) \Delta\eta ~\mathrm{Im} \left[ \zeta_{k_1}(\tau_e) \zeta_{k_2}(\tau_e) \zeta_{k_3}(\tau_e) \zeta_{\tilde{k}_1}^*(\tau_s) \zeta_{\tilde{k}_2}^*(\tau_s) \zeta_{\tilde{k}_3}'^*(\tau_s) \right] + \mathrm{perm.},
\end{align}
where factor $1/2$ arises because only the left half of the Dirac-Delta function is integrated. In the squeezed limit, it becomes
\begin{equation}
\left\langle \! \left\langle \bz_{\bk_1}(\tau_e) \bz_{\bk_2}(\tau_e) \bz_{-\bk_2}(\tau_e) \right\rangle \! \right\rangle = - \left\lbrace \frac{1}{2} \Delta\eta - 4 \Delta\eta \mpl^2 \epsilon(\tau_s) a^2(\tau_s) \mathrm{Im} \left[ \frac{\zeta_{k_2}^2(\tau_e)}{\abs{\zeta_{k_2}(\tau_e)}^2} \zeta_{k_2}^*(\tau_s) \zeta_{k_2}'^*(\tau_s) \right] \right\rbrace \abs{\zeta_{k_1}(\tau_e)}^2 \abs{\zeta_{k_2}(\tau_e)}^2. \label{bend}
\end{equation}
After adding contribution from field redefinition, the bispectrum of $\zeta$ in squeezed limit is
\begin{align}
\left\langle \! \left\langle \zeta_{\bk_1}(\tau_e) \zeta_{\bk_2}(\tau_e) \zeta_{-\bk_2}(\tau_e) \right\rangle \! \right\rangle = & \left\langle \! \left\langle \bz_{\bk_1}(\tau_e) \bz_{\bk_2}(\tau_e) \bz_{-\bk_2}(\tau_e) \right\rangle \! \right\rangle + \frac{\eta(\tau_e)}{4} 4 \left\langle \! \left\langle \bz_{\bk_1}(\tau_e) \bz_{-\bk_1}(\tau_e) \right\rangle \! \right\rangle \left\langle \! \left\langle \bz_{\bk_2}(\tau_e) \bz_{-\bk_2}(\tau_e) \right\rangle \! \right\rangle \nonumber\\
& + k_e^{-1} \left\langle \! \left\langle \bz_{\bk_1}(\tau_e) \bz_{-\bk_1}(\tau_e) \right\rangle \! \right\rangle \left( \langle \! \langle \bz_{\bk_2}'(\tau_e) \bz_{-\bk_2}(\tau_e) \rangle \! \rangle + \langle \! \langle \bz_{\bk_2}(\tau_e) \bz_{-\bk_2}'(\tau_e) \rangle \! \rangle \right),
\end{align}
and more explicitly
\begin{align}
\left\langle \! \left\langle \zeta_{\bk_1}(\tau_e) \zeta_{\bk_2}(\tau_e) \zeta_{-\bk_2}(\tau_e) \right\rangle \! \right\rangle = - &\left\lbrace \frac{1}{2} \Delta\eta - 4 \Delta\eta \mpl^2 \epsilon(\tau_s) a^2(\tau_s) \mathrm{Im} \left[ \frac{\zeta_{k_2}^2(\tau_e)}{\abs{\zeta_{k_2}(\tau_e)}^2} \zeta_{k_2}^*(\tau_s) \zeta_{k_2}'^*(\tau_s) \right] \right.  \nonumber\\
& \left. - 2 \mathrm{Re} \left[ \frac{\zeta_{k_2}(\tau_e) \zeta_{k_2}'^*(\tau_e)}{k_e \abs{\zeta_{k_2}(\tau_e)}^2} \right] - \eta(\tau_e)  \right\rbrace \abs{\zeta_{k_1}(\tau_e)}^2 \abs{\zeta_{k_2}(\tau_e)}^2. \label{bispec}
\end{align}
We need to define $\eta(\tau_e) = - \Delta\eta/2$ so the squeezed limit of bispectrum becomes a continuous function from $\tau_e^-$ to $\tau_e^+$. Therefore, bispectrums \eqref{bispec-}, \eqref{bispec+}, and \eqref{bispec} are equal.

\subsection{At the end of inflation \label{end}}
\begin{figure}[tbp]
\centering 
\includegraphics[width=0.7\textwidth]{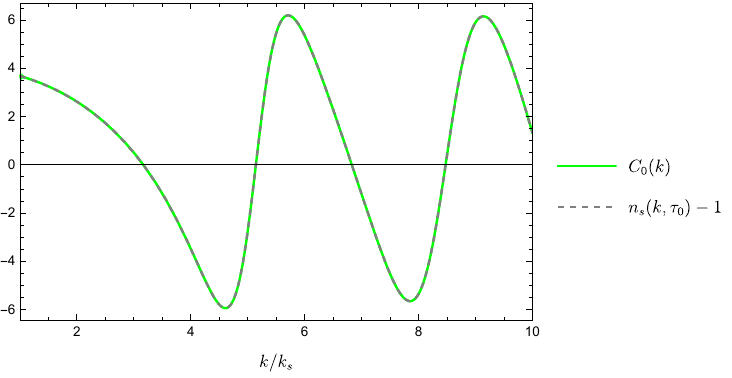}
\caption{\label{fig4} Plot of $C_0(k)$ and $n_s(k, \tau_0) - 1$. We choose $k_e/k_s = 10$ just for illustrative purposes.}
\end{figure}

At the end of inflation, $\tau_0 ~(\rightarrow 0)$, contribution from field redefinition are negligible because $\eta(\tau_0) \approx 0$ and $\dot{\zeta}$ decays. Therefore, the bispectrum of $\zeta$ and $\bz$ are equal
\begin{equation}
\langle \zeta_{\bk_1}(\tau_0) \zeta_{\bk_2}(\tau_0) \zeta_{\bk_3}(\tau_0) \rangle = \langle \bz_{\bk_1}(\tau_0) \bz_{\bk_2}(\tau_0) \bz_{\bk_3}(\tau_0) \rangle,
\end{equation}
and it is given by
\begin{equation}
\left\langle \! \left\langle \zeta_{\bk_1}(\tau_0) \zeta_{\bk_2}(\tau_0) \zeta_{\bk_3}(\tau_0) \right\rangle \! \right\rangle = -2 \mpl^2 \int_{-\infty}^{\tau_0} \md\tau_1 \epsilon(\tau_1) \eta'(\tau_1) a^2(\tau_1) \mathrm{Im} \left[ \zeta_{k_1}(\tau_0) \zeta_{k_2}(\tau_0) \zeta_{k_3}(\tau_0) \zeta_{\tilde{k}_1}^*(\tau_1) \zeta_{\tilde{k}_2}^*(\tau_1) \zeta_{\tilde{k}_3}'^*(\tau_1) \right] + \mathrm{perm}.
\end{equation}
Substituting \eqref{etadot} into the time integral yields,
\begin{align}
\left\langle \! \left\langle \zeta_{\bk_1}(\tau_0) \zeta_{\bk_2}(\tau_0) \zeta_{\bk_3}(\tau_0) \right\rangle \! \right\rangle = & - 2 \mpl^2 \epsilon(\tau_e)  a^2(\tau_e) \Delta\eta ~\mathrm{Im} \left[ \zeta_{k_1}(\tau_0) \zeta_{k_2}(\tau_0) \zeta_{k_3}(\tau_0) \zeta_{\tilde{k}_1}^*(\tau_e) \zeta_{\tilde{k}_2}^*(\tau_e) \zeta_{\tilde{k}_3}'^*(\tau_e) \right] \nonumber\\
& + 2 \mpl^2 \epsilon(\tau_s)  a^2(\tau_s) \Delta\eta ~\mathrm{Im} \left[ \zeta_{k_1}(\tau_0) \zeta_{k_2}(\tau_0) \zeta_{k_3}(\tau_0) \zeta_{\tilde{k}_1}^*(\tau_s) \zeta_{\tilde{k}_2}^*(\tau_s) \zeta_{\tilde{k}_3}'^*(\tau_s) \right] + \mathrm{perm}.
\end{align}
In the squeezed limit, the bispectrum becomes
\begin{align}
\left\langle \! \left\langle \zeta_{\bk_1}(\tau_0) \zeta_{\bk_2}(\tau_0) \zeta_{-\bk_2}(\tau_0) \right\rangle \! \right\rangle = - & \left\lbrace 4 \Delta\eta \mpl^2 \epsilon(\tau_e) a^2(\tau_e) \mathrm{Im} \left[ \frac{\zeta_{k_2}^2(\tau_0)}{\abs{\zeta_{k_2}(\tau_0)}^2} \zeta_{k_2}^*(\tau_e) \zeta_{k_2}'^*(\tau_e) \right] \right. \nonumber\\ 
& \left. - 4 \Delta\eta \mpl^2 \epsilon(\tau_s) a^2(\tau_s) \mathrm{Im} \left[ \frac{\zeta_{k_2}^2(\tau_0)}{\abs{\zeta_{k_2}(\tau_0)}^2} \zeta_{k_2}^*(\tau_s) \zeta_{k_2}'^*(\tau_s) \right] \right\rbrace \abs{\zeta_{k_1}(\tau_0)}^2 \abs{\zeta_{k_2}(\tau_0)}^2.
\end{align}
We define the coefficient as
\begin{equation}
C_0(k) = 4 \Delta\eta \mpl^2 \mathrm{Im} \left\lbrace \frac{\zeta_{k}^2(\tau_0)}{\abs{\zeta_{k}(\tau_0)}^2} \left[ \epsilon(\tau_e) a^2(\tau_e) \zeta_{k}^*(\tau_e) \zeta_{k}'^*(\tau_e) - \epsilon(\tau_s) a^2(\tau_s) \zeta_{k}^*(\tau_s) \zeta_{k}'^*(\tau_s) \right] \right\rbrace.
\end{equation}
The plot of $C_0(k)$ is shown in Fig. \ref{fig4}. For comparison, we also plot $n_s(k, \tau_0) - 1$ directly from power spectrum \eqref{ps0}. We can see that $C_0(k)$ precisely equals to $n_s(k, \tau_0) - 1$, which confirms Maldacena's theorem. The equality $C_0(k) = n_s(k, \tau_0) - 1$ holds exactly, which one can confirm by substituting mode function \eqref{zetasr2} to the explicit form of $C_0(k)$.

\section{One-loop correction \label{oneloop}}

Schematically, the one-loop correction to the power spectrum of $\zeta$ can be written as
\begin{equation}
\left\langle \! \left\langle \zeta_{\bp}(\tau_0) \zeta_{-\bp}(\tau_0) \right\rangle \! \right\rangle_{(1)} = \left\langle \! \left\langle \bz_{\bp}(\tau_0) \bz_{-\bp}(\tau_0) \right\rangle \! \right\rangle_{(1)} + \mathrm{redefinition ~terms}.
\end{equation}
The redefinition terms are negligible because they are evaluated at the end of SR period\footnote{Some time after this paper was posted on arXiv, Refs. \cite{Fumagalli:2023hpa, Tada:2023rgp} appeared, which calculated the one-loop correction in the same setup by using the third-order action with boundary terms. With different technical details, both papers claim cancellation between the one-loop correction induced by the bulk and boundary cubic self-interactions. In our approach, the boundary interactions and interactions proportional to the equation of motion are removed by field redefinition, as explained in Sec.~\ref{sec3}. At present, those papers are criticized by \cite{Firouzjahi:2023bkt}. We will solve this discrepancy elsewhere, pointing out what is missing in \cite{Fumagalli:2023hpa, Tada:2023rgp}.}. Thus, the one-loop correction at the end of inflation is simply
\begin{equation}
\left\langle \! \left\langle \zeta_{\bp}(\tau_0) \zeta_{-\bp}(\tau_0) \right\rangle \! \right\rangle_{(1)} = \left\langle \! \left\langle \bz_{\bp}(\tau_0) \bz_{-\bp}(\tau_0) \right\rangle \! \right\rangle_{(1)}.
\end{equation}
In this section, we calculate the one-loop correction to the large-scale power spectrum by two different methods: source method and direct in-in formalism. The source method utilizes bispectrum to calculate the one-loop correction, which is used by \cite{Riotto:2023hoz}. Prior to this work, source method is implemented by \cite{Syu:2019uwx, Cheng:2021lif}, in a similar context. On the other hand, direct in-in formalism is simply a second-order expansion of the in-in perturbation theory \eqref{inin}, as we did in \cite{Kristiano:2022maq}.

\subsection{Source method \label{source}}
Recall the second-order and third-order actions given by \eqref{S2}, \eqref{sbulk}, and \eqref{sbz}. The total second-order and leading bulk interaction reads
\begin{equation}
S[\bz] = S^{(2)}[\bz] + S_{\mathrm{bulk}}[\bz] = M_{\mathrm{pl}}^2 \int \mathrm{d}\tau ~\mathrm{d}^3x ~a^2 \epsilon  \left[ (\bz')^2 - (\partial_i \bz)^2 + \frac{1}{2} \eta' \bz' \bz^2 \right].
\end{equation}
The corresponding equation of motion for $\bz$ with long wavelength is
\begin{equation}
\bz_\bp'' + \frac{(a^2 \epsilon)'}{a^2 \epsilon} \bz_\bp' + \frac{(a^2 \epsilon \eta')'}{4 a^2 \epsilon} \int \frac{\md^3 k}{(2 \pi)^3} \bz_\bk \bz_{\bp-\bk} = 0,
\end{equation}
where $\bp$ is a wavevector on the CMB scale. The last term can be regarded as a source term of the second-order differential equation. The solution can be written as $\bz = \bz^f + \bz^s$, where $\bz^f$ and $\bz^s$ are the homogeneous and inhomogeneous solutions, respectively. The mode function of the homogeneous solution is
\begin{equation}
\bz^f_p(\tau) = \mathcal{A}_p + \mathcal{B}_p \int^\tau \frac{\md \tau_1}{a^2(\tau_1) \epsilon(\tau_1)},
\end{equation}
where $\mathcal{A}_p$ and $\mathcal{B}_p$ are arbitrary functions of $p$. The inhomogeneous solution is given by
\begin{equation}
\bz^s_\bp(\tau) = - \frac{1}{4} \int_{-\infty}^{\tau} \frac{\md \tau_1}{a^2(\tau_1) \epsilon(\tau_1)} \int_{-\infty}^{\tau_1} \md \tau_2 \left[ a^2(\tau_2) \epsilon(\tau_2) \eta'(\tau_2) \right]' \int \frac{\md^3 k}{(2 \pi)^3} \bz_\bk(\tau_2) \bz_{\bp-\bk}(\tau_2),
\end{equation}
then performing integration by parts leads to
\begin{align}
\bz^s_\bp(\tau_0) = & - \frac{1}{4} \int_{-\infty}^{\tau_0} \frac{\md \tau_1}{a^2(\tau_1) \epsilon(\tau_1)} a^2(\tau_1) \epsilon(\tau_1) \eta'(\tau_1) \int \frac{\md^3 k}{(2 \pi)^3} \bz_\bk(\tau_1) \bz_{\bp-\bk}(\tau_1) \nonumber\\
& + \frac{1}{4} \int_{-\infty}^{\tau_0} \frac{\md \tau_1}{a^2(\tau_1) \epsilon(\tau_1)} \int_{-\infty}^{\tau_1} \md \tau_2 ~ a^2(\tau_2) \epsilon(\tau_2) \eta'(\tau_2) \int \frac{\md^3 k}{(2 \pi)^3} \frac{\md}{\md \tau_2} \left[ \bz_\bk(\tau_2) \bz_{\bp-\bk}(\tau_2) \right].
\end{align}
Because we are interested in the finite effect of the amplified perturbation on small scale due to the USR period, the wavenumber integration domain is restricted to $k_s \lesssim k \lesssim k_e$. Substituting \eqref{etadot} to the time integral and use approximation $p \ll k$ yields \footnote{Note that the second term in \eqref{zetasp} cannot be neglected. The approximation $\zeta'_p(\tau_e) \ll aH \zeta_p(\tau_e)$ holds only for mode functions with $p \ll k_e$, which are far outside the horizon at $\tau_e$.}
\begin{equation}
\bz^s_\bp(\tau_0) = - \frac{1}{4} \Delta\eta \int \frac{\md^3 k}{(2 \pi)^3} \bz_\bk(\tau_e) \bz_{\bp-\bk}(\tau_e) + \frac{1}{4} \Delta\eta \int \frac{\md^3 k}{(2 \pi)^3} \frac{2}{3 k_e} \bz'_\bk(\tau_e) \bz_{\bp-\bk}(\tau_e). \label{zetasp}
\end{equation}
Note that each $\bz$ in this solution is an operator. 

The two-point function of $\bz$ can be written as
\begin{equation}
\left\langle \! \left\langle \bz_{\bp}(\tau_0) \bz_{-\bp}(\tau_0) \right\rangle \! \right\rangle = \langle \! \langle \bz^f_{\bp}(\tau_0) \bz^f_{-\bp}(\tau_0) \rangle \! \rangle + 2 \langle \! \langle \bz^s_{\bp}(\tau_0) \bz^f_{-\bp}(\tau_0) \rangle \! \rangle + \langle \! \langle \bz^s_{\bp}(\tau_0) \bz^s_{-\bp}(\tau_0) \rangle \! \rangle.
\end{equation}
The first term is simply the tree-level contribution 
\begin{equation}
\left\langle \! \left\langle \bz_{\bp}(\tau_0) \bz_{-\bp}(\tau_0) \right\rangle \! \right\rangle_{(0)} \equiv \langle \! \langle \bz^f_{\bp}(\tau_0) \bz^f_{-\bp}(\tau_0) \rangle \! \rangle = \abs{\zeta_p (\tau_0)}^2 ,
\end{equation}
with power spectrum given in \eqref{pssr}. The second and third terms are the one-loop corrections to the two-point function
\begin{equation}
\left\langle \! \left\langle \bz_{\bp}(\tau_0) \bz_{-\bp}(\tau_0) \right\rangle \! \right\rangle_{(1)} \equiv 2 \langle \! \langle \bz^s_{\bp}(\tau_0) \bz^f_{-\bp}(\tau_0) \rangle \! \rangle + \langle \! \langle \bz^s_{\bp}(\tau_0) \bz^s_{-\bp}(\tau_0) \rangle \! \rangle. \label{onelsfcor}
\end{equation}
The one-loop correction comes from the correlation between inhomogeneous and homogeneous solutions and the correlation of two inhomogeneous solutions.

First, we calculate the correlation of two inhomogeneous solutions. It reads
\begin{equation}
\langle \! \langle \bz^s_{\bp}(\tau_0) \bz^s_{-\bp}(\tau_0) \rangle \! \rangle = \left( \frac{\Delta\eta}{4} \right)^2 \int \frac{\md^3 k_1 \md^3 k_2}{(2 \pi)^6} \left\langle \! \left\langle \bar{\bz}_{\bk_1}(\tau_e) \bz_{\bp-\bk_1}(\tau_e) \bar{\bz}_{\bk_2}(\tau_e) \bz_{-\bp-\bk_2}(\tau_e) \right\rangle \! \right\rangle,
\end{equation}
where $\bar{\bz}$ is defined as
\begin{equation}
\bar{\bz}_{\bk}(\tau_e) = \bz_{\bk}(\tau_e) - \frac{2}{3 k_e} \bz'_{\bk}(\tau_e).
\end{equation}
Performing the Wick contraction, it becomes
\begin{equation}
\langle \! \langle \bz^s_{\bp}(\tau_0) \bz^s_{-\bp}(\tau_0) \rangle \! \rangle = \left( \frac{\Delta\eta}{4} \right)^2 \int \frac{\md^3 k}{(2 \pi)^3} \left[ \abs{\zeta_k - \frac{2}{3 k_e} \zeta'_k}^2 \abs{\zeta_k}^2 + \abs{\left(\zeta_k - \frac{2}{3 k_e} \zeta'_k \right) \zeta_k^*}^2 \right]_{\tau = \tau_e}.
\end{equation}
We can estimate how large is such correction. Performing integration around $k \sim k_s$ leads to
\begin{equation}
\langle \! \langle \bz^s_{\bp}(\tau_0) \bz^s_{-\bp}(\tau_0) \rangle \! \rangle \sim \mathcal{O}(1) \abs{\zeta_p (\tau_0)}^2 \frac{\abs{\zeta_{k_s} (\tau_e)}^2}{\abs{\zeta_p (\tau_0)}^2} \Delta_{s(\mathrm{PBH})}^{2}.
\end{equation}
Substituting typical numerical values for PBH formation, $p/k_s \sim 10^{-6}$ for PBH with mass $\mathcal{O}(10) M_\odot$, $\Delta_{s(\mathrm{PBH})}^{2} \sim 0.01$, and $\Delta_{s(\mathrm{SR})}^{2}(p) \sim 10^{-9}$, we obtain
\begin{equation}
\frac{\langle \! \langle \bz^s_{\bp}(\tau_0) \bz^s_{-\bp}(\tau_0) \rangle \! \rangle}{\left\langle \! \left\langle \bz_{\bp}(\tau_0) \bz_{-\bp}(\tau_0) \right\rangle \! \right\rangle_{(0)}} \sim \mathcal{O}(1) \frac{\left[ \Delta_{s(\mathrm{PBH})}^{2} \right]^2}{\Delta_{s(\mathrm{SR})}^{2}(p)} \left( \frac{p}{k_s} \right)^3 \ll 1.
\end{equation}
Therefore, the correlation of two inhomogeneous solutions is very small because of cubic suppression between large and small scale.

Next, we calculate the correlation between inhomogeneous and homogeneous solutions. It reads
\begin{equation}
\langle \! \langle \bz^s_{\bp}(\tau_0) \bz^f_{-\bp}(\tau_0) \rangle \! \rangle = \frac{1}{4} \Delta\eta \int \frac{\md^3 k}{(2 \pi)^3} \left[ - \langle \! \langle \bz_\bk(\tau_e) \bz_{-\bk}(\tau_e) \bz_{\bp}(\tau_e) \rangle \! \rangle + \frac{2}{3 k_e} \langle \! \langle \bz'_\bk(\tau_e) \bz_{-\bk}(\tau_e) \bz_{\bp}(\tau_e) \rangle \! \rangle  \right]. \label{sfcor}
\end{equation}
Note that we use $\bz_{\bp}(\tau_0) \approx \bz_{\bp}(\tau_e)$. This can be understood intutively as follows. We are calculating the one-loop correction generated by cubic self-interaction. The vertex factor of the one-loop correction corresponds to the bispectrum. Such a bispectrum evolves in time with the dominant contribution at $\tau = \tau_e$. The one-loop correction at the end of inflation is obtained by integrating the bispectrum over time to the end of inflation, which captures the main contribution at $\tau = \tau_e$. 

In this source method, the one-loop correction to the large-scale power spectrum is proportional to the squeezed limit of the three-point functions. The first term is nothing but the squeezed limit of the bispectrum that is given by \eqref{bend}. The second term are three-point correlations involving the time derivative of $\bz$, which can be calculated from in-in perturbation theory \eqref{inin1}. The squeezed limit of such correlations, when one $\bz$ is a long-wavelength perturbation, is given by
\begin{equation}
\langle \! \langle \bz'_{\bk}(\tau_e) \bz_{-\bk}(\tau_e) \bz_{\bp}(\tau_e) \rangle \! \rangle = -2 \mpl^2 \int_{-\infty}^{\tau_e} \md\tau_1 \epsilon(\tau_1) \eta'(\tau_1) a^2(\tau_1)  \mathrm{Im} \left[ \zeta'_{k}(\tau_e) \zeta_{k}(\tau_e) \zeta_{p}(\tau_e) \zeta_{\tilde{k}}'^*(\tau_1) \zeta_{\tilde{k}}^*(\tau_1) \zeta_{\tilde{p}}^*(\tau_1) \right] + \mathrm{perm}.
\end{equation}
Note that permutated terms with $\zeta_p'$ can be neglected because they are much smaller than the others. Performing a time integral with $\eta'(\tau)$ given in \eqref{etadot}, we obtain
\begin{align}
\langle \! \langle \bz'_{\bk}(\tau_e) \bz_{-\bk}(\tau_e) \bz_{\bp}(\tau_e) \rangle \! \rangle = & -4 \mpl^2 \epsilon(\tau_e) a^2(\tau_e) \frac{1}{2} \Delta\eta  \abs{\zeta_{p}(\tau_e)}^2 \abs{\zeta_{k}(\tau_e)}^2 \mathrm{Im}(\zeta_{k}' (\tau_e) \zeta_{k}'^*(\tau_e)) \nonumber\\
& + 4 \mpl^2 \epsilon(\tau_s) a^2(\tau_s) \Delta\eta \abs{\zeta_{p}(\tau_e)}^2 \mathrm{Im} \left[ \zeta'_{k}(\tau_e) \zeta_{k}(\tau_e) \zeta_{k}'^*(\tau_s) \zeta_{k}^*(\tau_s) \right]. \label{cordot}
\end{align}
The first term obviously vanishes, so only the second term contributes to the correlations. Substituting \eqref{bend} and \eqref{cordot} into \eqref{sfcor} yields \footnote{Compared with \cite{Riotto:2023hoz}, the author obtains only the first term in \eqref{sfcor}, although there is the second term due to integration by parts. However, this is not an essential difference. A more important discrepancy is \cite{Riotto:2023hoz} substitutes Maldacena's theorem to the first term, the squeezed bispectrum. The bispectrum of $\bz$ at the end of the USR period does not satisfy Maldacena's theorem, as explained in Sec.~\ref{usr1} to Sec.~\ref{usr3}. The bispectrum of $\bz$ at the end of inflation satisfies Maldacena's theorem because it is equal to the bispectrum of $\zeta$, as explained in Sec.~~\ref{end}. The one-loop correction to the power spectrum at the end of inflation is proportional to the bispectrum of $\bz$ at the end of the USR period, not the bispectrum of $\bz$ at the end of inflation. Therefore, substituting Maldacena's theorem to the first term into \eqref{sfcor} is incorrect. \label{foot1}}
\begin{align}
\left\langle \! \left\langle \zeta_{\bp}(\tau_0) \zeta_{-\bp}(\tau_0) \right\rangle \! \right\rangle_{(1)} = & ~\frac{1}{2} (\Delta\eta)^2 \abs{\zeta_{p}(\tau_e)}^2   \label{onels} \\
& \times\int \frac{\md^3 k}{(2 \pi)^3} \left\lbrace \frac{1}{2} \abs{\zeta_{k}(\tau_e)}^2 - 4 \mpl^2 \epsilon(\tau_s) a^2(\tau_s) \mathrm{Im} \left[ \left( \zeta_{k}(\tau_e) - \frac{2}{3k_e} \zeta'_{k}(\tau_e) \right) \zeta_{k}(\tau_e) \zeta_{k}^*(\tau_s) \zeta_{k}'^*(\tau_s) \right] \right\rbrace. \nonumber
\end{align}
We compare this result to direct in-in formalism in the next subsection.

\subsection{Direct in-in formalism \label{direct}}
The second-order expansion of in-in perturbation theory reads,
\begin{gather}
\langle \mathcal{O(\tau)} \rangle = \langle \mathcal{O(\tau)} \rangle_{(0,2)}^\dagger + \langle \mathcal{O(\tau)} \rangle_{(1,1)} + \langle \mathcal{O(\tau)} \rangle_{(0,2)}, \\
\langle \mathcal{O(\tau)} \rangle_{(1,1)} = \int_{-\infty}^{\tau} \mathrm{d}\tau_1 \int_{-\infty}^{\tau} \mathrm{d}\tau_2 \langle H_{\mathrm{int}}(\tau_1) \mathcal{\hat{O}} (\tau) H_{\mathrm{int}}(\tau_2) \rangle, \nonumber\\
\langle \mathcal{O(\tau)} \rangle_{(0,2)} = - \int_{-\infty}^{\tau} \mathrm{d}\tau_1 \int_{-\infty}^{\tau_1} \mathrm{d}\tau_2 \langle \mathcal{\hat{O}} (\tau) H_{\mathrm{int}}(\tau_1) H_{\mathrm{int}}(\tau_2) \rangle. \nonumber
\end{gather}
Because we are interested in the one-loop correction to the large-scale power spectrum, the operator $\mathcal{O(\tau)}$ is $\bz_\bp(\tau_0) \bz_{-\bp}(\tau_0)$. Substituting the leading interaction \eqref{hintl} leads to,
\begin{gather}
\langle \bz_{\bp}(\tau_0) \bz_{-\bp}(\tau_0) \rangle_{(1,1)} = \frac{1}{4} \mpl^4 \int_{-\infty}^{\tau_0} \md \tau_1 ~a^2(\tau_1) \epsilon(\tau_1) \eta'(\tau_1) \int_{-\infty}^{\tau_0} \md \tau_2 ~a^2(\tau_2) \epsilon(\tau_2) \eta'(\tau_2)  \nonumber \\
\int \prod_{a = 1}^6 \left[ \frac{\md^3 k_a}{(2\pi)^3} \right] \delta(\bk_1+\bk_2+\bk_3) \delta(\bk_4+\bk_5+\bk_6) \left\langle \bz_{\bk_1}'(\tau_1) \bz_{\bk_2}(\tau_1) \bz_{\bk_3}(\tau_1) \bz_{\bp}(\tau_0) \bz_{-\bp}(\tau_0) \bz_{\bk_4}'(\tau_2) \bz_{\bk_5}(\tau_2) \bz_{\bk_6}(\tau_2) \right\rangle, \\
\langle \bz_{\bp}(\tau_0) \bz_{-\bp}(\tau_0) \rangle_{(0,2)} = -\frac{1}{4} \mpl^4 \int_{-\infty}^{\tau_0} \md \tau_1 ~a^2(\tau_1) \epsilon(\tau_1) \eta'(\tau_1) \int_{-\infty}^{\tau_1} \md \tau_2 ~a^2(\tau_2) \epsilon(\tau_2) \eta'(\tau_2) \nonumber \\
\int \prod_{a = 1}^6 \left[ \frac{\md^3 k_a}{(2\pi)^3} \right] \delta(\bk_1+\bk_2+\bk_3) \delta(\bk_4+\bk_5+\bk_6) \left\langle \bz_{\bp}(\tau_0) \bz_{-\bp}(\tau_0)  \bz_{\bk_1}'(\tau_1) \bz_{\bk_2}(\tau_1) \bz_{\bk_3}(\tau_1) \bz_{\bk_4}'(\tau_2) \bz_{\bk_5}(\tau_2) \bz_{\bk_6}(\tau_2) \right\rangle.
\end{gather}
The total one-loop correction reads,
\begin{equation}
\left\langle \! \left\langle \bz_{\bp}(\tau_0) \bz_{-\bp}(\tau_0) \right\rangle \! \right\rangle_{(1)} = \left\langle \! \left\langle \bz_{\bp}(\tau_0) \bz_{-\bp}(\tau_0) \right\rangle \! \right\rangle_{(1,1)} + 2 \mathrm{Re} \left\langle \! \left\langle \bz_{\bp}(\tau_0) \bz_{-\bp}(\tau_0) \right\rangle \! \right\rangle_{(0,2)}.
\end{equation}
Performing the time integral with $\eta'(\tau)$ given in \eqref{etadot} and the Wick contraction, we obtain
\begin{align}
&\left\langle \! \left\langle \bz_{\bp}(\tau_0) \bz_{-\bp}(\tau_0) \right\rangle \! \right\rangle_{(1)} = \frac{1}{4} \mpl^4 (\Delta\eta)^2 \abs{\zeta_p(\tau_0)}^2 \int \frac{\md^3 k}{(2\pi)^3} 16 \left\lbrace  \left[\epsilon^2  a^4\abs{\zeta_k}^2 \mathrm{Im}(\zeta_p \zeta_p'^*) \mathrm{Im}(\zeta_k \zeta_k'^*) \right]_{\tau = \tau_e}  \right. \nonumber\\
&- 4 \epsilon(\tau_e) a^2(\tau_e) \epsilon(\tau_s) a^2(\tau_s) \mathrm{Im}(\zeta_p(\tau_0) \zeta_p^*(\tau_e)) \mathrm{Im} \left( \zeta'_k(\tau_e) \zeta_k(\tau_e) \zeta_{k}^*(\tau_s) \zeta_{k}'^*(\tau_s) \right) \nonumber \\
&\left. - 2 \epsilon(\tau_e) a^2(\tau_e) \epsilon(\tau_s) a^2(\tau_s) \mathrm{Im}(\zeta_p(\tau_e) \zeta_p'^*(\tau_e)) \mathrm{Im} \left( \zeta_{k}^2(\tau_e) \zeta_{k}^*(\tau_s) \zeta_{k}'^*(\tau_s) \right) + \left[\epsilon^2  a^4\abs{\zeta_k}^2 \mathrm{Im}(\zeta_p \zeta_p'^*) \mathrm{Im}(\zeta_k \zeta_k'^*)  \right]_{\tau = \tau_s} \right\rbrace.  \label{onelall}
\end{align}
The last term is much smaller than the other terms because the curvature perturbation is not amplified at $\tau = \tau_s$ yet. After some algebra, we find, \footnote{Note that we have to consider the difference between $\zeta_p(\tau_e)$ and $\zeta_p(\tau_0)$ for the second term in \eqref{onelall} because it is comparable to $\mathrm{Im}(\zeta_p(\tau_e) \zeta_p'^*(\tau_e))$ in the first and third term. From \eqref{zetasr2}, we can obtain
\begin{equation}
\mathrm{Im}(\zeta_p(\tau_0) \zeta_p^*(\tau_e)) = - \left( \frac{H^2}{4 M_{\mathrm{pl}}^2 \epsilon_{\mathrm{SR}} p^3} \right)_\star \frac{p^3}{3 k_s^3} \left( \frac{k_e}{k_s} \right)^3. \label{imzeta2}
\end{equation}
Substituting \eqref{imzeta} and \eqref{imzeta2} to \eqref{onelall} leads to \eqref{oneld}.
}
\begin{align}
\left\langle \! \left\langle \zeta_{\bp}(\tau_0) \zeta_{-\bp}(\tau_0) \right\rangle \! \right\rangle_{(1)} = & ~\frac{1}{4} (\Delta\eta)^2 \abs{\zeta_{p}(\tau_e)}^2 \label{oneld} \\
& \times \int \frac{\md^3 k}{(2 \pi)^3} \left\lbrace \abs{\zeta_{k}(\tau_e)}^2 - 8 \mpl^2 \epsilon(\tau_s) a^2(\tau_s) \mathrm{Im} \left[ \left( \zeta_{k}(\tau_e) - \frac{2}{3k_e} \zeta'_{k}(\tau_e) \right) \zeta_{k}(\tau_e) \zeta_{k}^*(\tau_s) \zeta_{k}'^*(\tau_s) \right] \right\rbrace. \nonumber
\end{align}
Hence, the calculation of the one-loop correction by the direct in-in formalism and source method \eqref{onels} leads to the same result.

\begin{figure}[tbp]
\centering 
\includegraphics[width=0.7\textwidth]{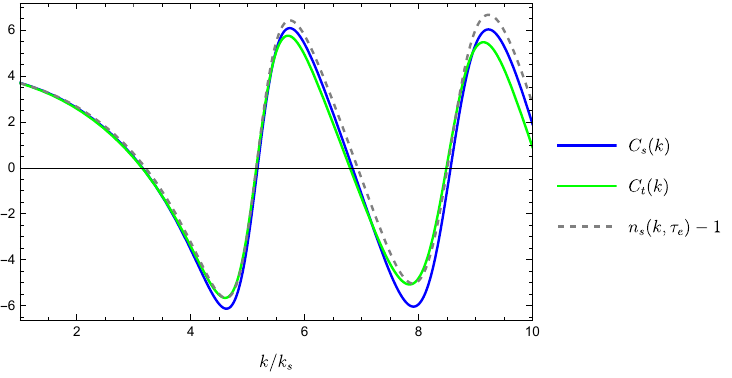}
\caption{\label{fig5} Plot of $C_s(k)$, $C_t(k)$, and $n_s(k, \tau_e) - 1$. We choose $k_e/k_s = 10$ just for illustrative purposes.}
\end{figure}

One-loop correction \eqref{oneld} can be written as \footnote{Incorrectly substituting Maldacena's theorem to \eqref{sfcor} leads to subtraction of the second term in \eqref{onelin} by the contribution from the field redefinition at the end of USR period. Recall equality $n_s(k, \tau_e) - 1 = \Delta\eta - B(k) + C_s(k)$ that we discussed below Eq. \eqref{bk}. From Fig. \ref{fig3}, we can see that $B(k) \approx \Delta\eta$. Substituting Maldacena's theorem means that one implicitly includes the contribution from the field redefinition $B(k)$, which almost cancels $\Delta\eta$. This is the reason Ref. \cite{Riotto:2023hoz} obtains only the first term in \eqref{onelin}, although there is a factor of $2$ discrepancy that might come from the prefactor of \eqref{onelsfcor}. \label{foot2}}
\begin{equation}
\left\langle \! \left\langle \zeta_{\bp}(\tau_0) \zeta_{-\bp}(\tau_0) \right\rangle \! \right\rangle_{(1)} = \Delta\eta \abs{\zeta_{p}(\tau_e)}^2 \int \frac{\md^3 k}{(2 \pi)^3} \left[  \frac{1}{2} C_s(k) - C_t(k) + \frac{1}{4} \Delta\eta \right] \abs{\zeta_{k}(\tau_e)}^2 , \label{oneldc}
\end{equation}
where $C_s(k)$ is defined in \eqref{cecs} and $C_t(k)$ is defined as
\begin{equation}
C_t(k) =  - 4 \Delta\eta \mpl^2 \epsilon(\tau_s) a^2(\tau_s)  \mathrm{Im} \left[ \frac{ \zeta'_k(\tau_e) \zeta_k(\tau_e) }{3 k_e \abs{\zeta_{k}(\tau_e)}^2} \zeta_{k}^*(\tau_s) \zeta_{k}'^*(\tau_s) \right].
\end{equation}
Plot of $C_t(k)$ is shown in Fig. \ref{fig5}. For comparison, we also plot $C_s(k)$ and $n_s(k,\tau_e) - 1$ in the same figure. We can see that $C_s(k) \approx C_t(k) \approx n_s(k, \tau_e) - 1$, so we can approximate the integral as
\begin{align}
\int \frac{\md^3 k}{(2 \pi)^3} \left[  \frac{1}{2} C_s(k) - C_t(k) \right] \abs{\zeta_{k}(\tau_e)}^2 & \approx - \frac{1}{2} \int \frac{\md^3 k}{(2 \pi)^3} (n_s(k, \tau_e) - 1) \abs{\zeta_{k}(\tau_e)}^2 \nonumber\\
& = - \frac{1}{2} \int \md \log k ~\Delta_s^2(k, \tau_e)  \frac{\md \log \Delta_s^2(k, \tau_e)}{\md \log k} \approx - \frac{1}{2} \Delta_{s(\mathrm{PBH})}^{2}.
\end{align}
Then, the one-loop correction becomes
\begin{equation}
\left\langle \! \left\langle \zeta_{\bp}(\tau_0) \zeta_{-\bp}(\tau_0) \right\rangle \! \right\rangle_{(1)} \approx - \frac{1}{2} \Delta\eta \abs{\zeta_{p}(\tau_e)}^2 \Delta_{s(\mathrm{PBH})}^{2} + \frac{1}{4} (\Delta\eta)^2 \abs{\zeta_{p}(\tau_e)}^2 \int \frac{\md^3 k}{(2 \pi)^3} \abs{\zeta_{k}(\tau_e)}^2. \label{onelin}
\end{equation}
We compare the one-loop correction to the tree-level contribution as
\begin{equation}
\frac{\left\langle \! \left\langle \zeta_{\bp}(\tau_0) \zeta_{-\bp}(\tau_0) \right\rangle \! \right\rangle_{(1)}}{\left\langle \! \left\langle \zeta_{\bp}(\tau_0) \zeta_{-\bp}(\tau_0) \right\rangle \! \right\rangle_{(0)}} \approx - \frac{1}{2} (\Delta\eta) \Delta_{s(\mathrm{PBH})}^{2} + \frac{1}{4} (\Delta\eta)^2 \int \frac{\md^3 k}{(2 \pi)^3} \abs{\zeta_{k}(\tau_e)}^2.
\end{equation}
In order to trust perturbation theory, this ratio must be much smaller than unity. The first term is the one-loop correction predicted by \cite{Riotto:2023hoz}. For $\Delta_{s(\mathrm{PBH})}^{2} \sim 0.1$, the ratio is $\mathcal{O}(0.1)$, so contribution of the first term is quite small \footnote{The reason can be intuitively understood as follows. Integral in \eqref{oneldc} can be written as
\begin{equation}
\int_{k_s}^{k_e} \frac{\md k}{k} ~C(k) \Delta_s^2(k, \tau_e),
\end{equation}
where $C(k)$ is a function coefficient. Because $\Delta_s^2(k, \tau_e) \approx \Delta_{s(\mathrm{PBH})}^{2} $ within the integration domain, the integral strongly depends on the coefficient. Integral with coefficient $C_s(k)$ or $C_t(k)$ will be significantly smaller than integral with constant coefficient, because $C_s(k)$ or $C_t(k)$ is oscillating around zero. The first term in \eqref{onelin} comes from integral with coefficient $C_s(k)$ and $C_t(k)$, while the second term comes from integral with constant coefficient $\Delta\eta$. Therefore, it is reasonable that the second term is dominant compared to the first term.
}. However, there is also the second term. Requiring the second term to be much smaller than unity leads to the upper bound $\Delta_{s(\mathrm{PBH})}^{2} \ll \mathcal{O}(0.01)$. 

In Eq. \eqref{onelin}, we obtain the bare one-loop correction. Issue related to regularization and renormalization is discussed in \cite{Kristiano:2022maq}. Although we focus on SR to USR to SR transition with $\Delta\eta \approx 6$, our result is general for any sharp $\eta$ transition. We briefly discuss other PBH formation models from single-field inflation in \cite{Kristiano:2022maq}.

\section{Conclusion \label{sec5}}
In this paper, we consider a single-field inflation model with sharp transition of the second SR parameter from SR to USR to SR period, which leads to PBH formation. We have derived the bispectrum and one-loop correction in such a setup. We have shown explicitly that the bispectrum in squeezed limit satisfies Maldacena's theorem. We have also clarified which terms in the cubic self-interaction of the curvature perturbation yield contribution to the squeezed bispectrum. At the end of the USR period, the bispectrum is generated by the bulk interaction and field redefinition. At the end of inflation, bispectrum is generated only by bulk interaction. We have also demonstrated that the calculation of the one-loop correction by source method and direct in-in formalism leads to the same result, confirming the conclusion of our paper \cite{Kristiano:2022maq} that the one-loop correction to the large-scale power spectrum provides a significant constraint to the small-scale power spectrum.

\acknowledgements

J.~K. acknowledges the support from JSPS KAKENHI Grants No.~22KJ1006 and No.~22J20289. J.~K. is also supported by the Global Science Graduate Course (GSGC) program of The University of Tokyo. J.~Y. is supported by JSPS KAKENHI Grant No.~20H05639 and Grant-in-Aid for Scientific Research on Innovative Areas 20H05248. 

\bibliographystyle{apsrev4-1}
\bibliography{reference}

\end{document}